\documentclass[aps,superscriptaddress,nofootinbib,eqsecnum,notitlepage,twocolumn,prc,showpacs]{revtex4} 

\usepackage{amssymb}   
\usepackage{amsmath}   
\usepackage{graphicx}  
\usepackage{graphicx}
\usepackage{subfigure}
\usepackage{hyperref}
\usepackage{float}
\usepackage{color}
\usepackage{slashed}
\usepackage{multirow}

\usepackage{calrsfs}
\usepackage{urwchancal}
\DeclareMathAlphabet{\mathpzc}{OT1}{pzc}{m}{it}
\DeclareMathAlphabet{\pazocal}{OMS}{zplm}{m}{n}

\begin{document}

%
%
%

\title{Anisotropy in the equation of state of strongly magnetized quark matter within the Nambu--Jona-Lasinio model Model}

\author{ Sidney S. Avancini}
\affiliation{Departamento de F\'{\i}sica, Universidade Federal de Santa Catarina, 
Florian\'{o}polis, SC 88040-900, Brazil}
\email{sidney.avancini@ufsc.br}

\author{Veronica Dexheimer}
\affiliation{Department of Physics, Kent State University, Kent OH 44242 USA}
\email{vdexheim@kent.edu}

\author{Ricardo L. S. Farias}
\affiliation{Departamento de F\'{\i}sica, Universidade Federal de Santa Maria, 
Santa Maria, RS 97105-900, Brazil}
\email{ricardo.farias@ufsm.br}

\author{Varese S. Tim\'oteo}
\affiliation{Grupo de \'Optica e Modelagem Num\'erica - GOMNI, Faculdade de Tecnologia - FT, \\
Universidade Estadual de Campinas - UNICAMP, 13484-332 Limeira, SP , Brazil}
\email{varese@ft.unicamp.br}

\begin{abstract}
In this article, we calculate the magnetization and other thermodynamical quantities for strongly 
magnetized quark matter 
within the Nambu-Jona-Lasinio model at zero temperature. We assume two scenarios, chemically 
equilibrated charge
neutral matter present in the interior of compact stars and zero-strangeness isospin-symmetric 
matter created in nuclear experiments. We show that the magnetization 
oscillates with density
but in a much more smooth form than what was previously shown in the literature. As a consequence, we do not see the 
unphysical behavior in the pressure in the direction perpendicular to the magnetic field that was previously found. 
Finally, we
also analyze the effects of a vector interaction on our results.
\end{abstract}


\maketitle

\section{Introduction}

Understanding dense and/or hot matter in the presence of strong magnetic fields is one of the most 
important challenges of
nuclear physics today. At low chemical potentials, extremely high magnetic fields have been estimated 
to be breafly created 
in relativistic heavy-ion collisions \cite{PhysRevD.78.074033,PhysRevD.80.034028,KHARZEEV2009543c,
doi:10.1142/S0217751X09047570,PhysRevC.83.054911}, with strengths of up to $10^{19}$ and $10^{20}$ G expected to be 
generated during 
non-central heavy-ion collisions at the BNL Relativistic Heavy Ion Collider (RHIC) and at European Organization for Nuclear Research (CERN), respectively. In this regime, the role played 
by magnetic fields in quark deconfinement and chiral symmetry restoration can, to some extent, be extracted from lattice 
QCD data. 

At low temperatures, high magnetic fields have been measured on the surface of neutron stars and 
extremely high magnetic
fields have been inferred to exist in their interior. More specifically, measurements using anharmonic 
precession of star 
spin down have estimated surface magnetic fields to be of the order of $10^{15}$ G for the 
sources 1E~1048.1-5937
and 1E~2259+586 \cite{Melatos:1999ji} and data for slow phase modulations in star hard x-ray 
pulsations (interpreted 
as free precession) suggest internal magnetic fields to be on the magnitude of $10^{16}$ G for the source
4U 0142+61 \cite{Makishima:2014dua}. Together, these estimates have motivated a large amount of research 
on the issue of how magnetic fields modify the microscopic structure (represented in the equation of state) and 
the macroscopic structure (obtained from the solution of the Einstein-Maxwell's equations) of neutron stars.
Unfortunately, in this regime, there is no guidance from lattice QCD concerning the effect of magnetic fields 
on deconfined quark matter.

At high enough baryonic chemical potential/density a deconfinement transition to quark matter takes place and, when the temperature is low enough, other more complex phases such as color superconducting phases or 
inhomogeneous chiral condensates become energetically favorable. Much effort has been made to understand the physics of these phases in the presence of strong magnetic fields~\cite{ferrer_notes}. The most favored phase of QCD at high densities is the color-flavor-looked (CFL) superconducting phase~\cite{alfordNPB} and, for a magnetic field strength of the order of the quark energy gap, a magnetic-CFL phase is preffered~\cite{2007PhRvD..76d5011F,2007AIPC..947..395D,2005PhRvL..95o2002F,2006NuPhB.747...88F,2006JPhA...39.6349F}. For field strengths comparable to the magnetic masses of charged gluons,  the formation of a gluon-vortex state can take place~\cite{2006PhRvL..97l2301F,2007JPhA...40.6913F} and, as explained in~\cite{2007PhRvD..76d5011F}, the vortex formation corresponds to a phase transition from a magnetic-CFL to a Paramagnetic-CFL phase. There are many other effects produced by a strong magnetic fields in combination with superconductivity~\cite{2007PhRvD..76k4012F,2007PhRvD..76j5030N,PhysRevLett.100.032007,2011NuPhB.853..213F,2011PhLB..706..232F,2012PhRvD..85j3529F}, such as the BEC-BCS crossover~\cite{2011PhRvD..84f5014W,2016PhRvD..93b5017D,PhysRevD.78.074033,Duarte:2017nzv,Duarte:2016pgi} and 
 the modification of	 chiral inhomogeneous phases~\cite{2010PhRvD..82g6002F,2015PhLB..743...66T,Ferrer:2012zq,2015PhRvD..92j5018C}.
 
For simplicity, in this article we make use of the Nambu-Jona-Lasinio (NJL) model without pairing to describe zero strangeness isospin-symmetric matter (such as that created in nuclear experiments) and chemically equilibrated charge neutral matter (such as that present in the interiors of compact stars) to study how magnetic fields 
influence cold quark matter. We find that, unlike what was previously stated in the literature for this version of the 
model~\cite{Menezes:2015jka,Menezes:2015mqa}, strong magnetic fields do not generate unphysical behavior of thermodynamical quantities, such as the magnetization and pressure in the direction perpendicular to the magnetic field. In addition, we verify that our conclusions hold even when vector interactions (which allow us to reproduce astrophysical constraints) are added to the model.

 \section{The Model}

The description of matter in this work includes three-flavored quark matter and leptons
(electrons and muons). While 
the leptons are described by a free Fermi gas under the influence of magnetic fields
(which includes the
quantization of Landau levels, see Ref.~\cite{Strickland:2012vu} and references 
therein for details), 
the description of strongly interacting quarks is much more complicated. The Lagrangian 
density for the SU(3) NJL 
model reads \cite {Nambu:1961fr,Nambu:1961tp,Buballa:2003qv}:
\begin{eqnarray}
{\pazocal L}_f = {\pazocal L}_{Dir}+ {\pazocal L}_{sym}+{\pazocal L}_{det} \ ,
\end{eqnarray}
where the different terms stand for the Dirac, symmetric
(four-point interaction) and t'Hooft (six-point interaction) terms:
\begin{eqnarray}
{\pazocal L}_{Dir}= {\bar{\psi}}_f \left[\gamma_\mu\left(i\partial^{\mu}
- {\hat q}_f A^{\mu} \right)-
{\hat m}_c \right ] \psi_f  \ ,
\end{eqnarray}
\begin{eqnarray}
{\pazocal L}_{sym}=G \sum_{a=0}^8 \left [({\bar \psi}_f \lambda_ a \psi_f)^2 + ({\bar \psi}_f i\gamma_5 \lambda_a
 \psi_f)^2 \right ]  \ ,
\end{eqnarray}
\begin{eqnarray}
{\pazocal L}_{det}&=&-K  \big\{ {\rm det}_f \left [ {\bar \psi}_f(1+\gamma_5) \psi_f \right] \nonumber\\
&+&  {\rm det}_f \left [ {\bar \psi}_f(1-\gamma_5) \psi_f \right] \big\} \ ,
\end{eqnarray}
where $\psi_f = (u,d,s)^T$ represents the three-flavored quark field,
while ${\hat m}_c= {\rm diag}_f (m_u,m_d,m_s)$ 
and ${\hat q}_f={\rm diag}(q_u,q_d,q_s)$ are the quark mass and charge matrices. The interaction
with the electromagnetic 
field appears in the Dirac term through the vector potential $A_\mu$. The coupling constants 
$G$ and $K$ are to be 
determined, $\lambda_0=\sqrt{2/3}I$  with $I$ being the unit matrix in the three flavor 
space, and $1<\lambda_a\le 8$
denote the Gell-Mann matrices. For the leptons, we use the mass values $m_e=0.511$ MeV and  $m_\mu=105.66$ MeV.

As the NJL model is non renormalizable, we apply a sharp ultraviolet cut-off $\Lambda$ in 
three-momentum space. The
parameters of the model, $\Lambda$, $G$ and $K$, and the current quark masses, $m_u=m_d$ and $m_s$, 
are determined  
by fitting $f_\pi$, $m_\pi$, $m_K$, and $m_{\xi'}$ to their empirical values. In this work, we adopt
the parametrization 
proposed in Ref.~\cite{Hatsuda:1994pi} with $\Lambda = 631.4$ MeV, 
$G =1.835/\Lambda^2$, $K=9.29/ \Lambda^5$, $m_u= m_d=5.5$ MeV, and $m_s=135.7$ MeV.

The thermodynamical potential for the quark sector at zero temperature reads:
\begin{eqnarray}
\Omega = -P = {\pazocal E} - \sum_f {\mu}_f {\rho}_f \ ,
\end{eqnarray} 
where $P$ (also referred to as $P_\parallel$) is the pressure in the direction of the magnetic field, ${\pazocal E}$ the energy density,  $\mu_f$ the quark flavor chemical
potential, and $\rho_f$ the 
quark flavor number density. A similar expression can be written for the leptonic sector. In the 
following results, 
normalization terms are implied in order to have $\Omega=0$
(and also $\Omega_{l}=0$ for the leptons)
when the quark and leptonic chemical potentials are set to zero.

It is important to stress that in this work, as a result of the normalization described above, we do not account for the pure electromagnetic contribution in the Lagrangian density or in other thermodynamical quantities. The pure electromagnetic contribution is independent of any matter contribution, unless one is concerned with equilibrium configurations of macroscopic properties of stars and is solving EinsteinÕs equations coupled to Maxwell's equations, which is not the case in this work. For detailed analyses that compare pure magnetic field contributions with the ones from magnetized matter inside individual stars, see Refs.~\cite{Chatterjee:2014qsa,Franzon:2015sya}.
\subsection{Number density}

In the mean field approximation, the quark number density for one quark flavor at zero temperature 
in the presence of
an external magnetic field with strength $B$ in one direction is simply:
\begin{eqnarray}
\rho_f=
\sum_{\nu=0}^{\nu_{f,max}} \alpha_\nu \frac{|q_f| B N_c }{2 \pi^2} p_{f}\ ,
\end{eqnarray}
where the sum over Landau levels $\nu$ goes until $\nu_{f, max} $ defined as the largest integer 
less than or equal 
to $({\mu_f^2 -M_f^2})/({2 |q_f|B})$. 
The degeneracy factor for each Landau level is $\alpha_0=1,\,\alpha_{k>0}=2$. $N_c$ is the 
number of colors and the
energy dispersion is given by $\mu_f= \sqrt{p_{f}^2 + s_f(\nu,B)^2} $, with $p_{f}$ being the 
Fermi momentum in the 
direction of the magnetic field and $s_f(\nu,B)= \sqrt {M_f^2 + 2 |q_f| B \nu}$ the quark effective 
mass modified 
by the magnetic field.

\subsection{Pressure}

In the mean field approximation, the quark thermodynamical pressure can be written as:
\begin{eqnarray}
P = \theta_u+\theta_d+\theta_s -2G(\phi_u^2+\phi_d^2+\phi_s^2) + 4K \phi_u \phi_d \phi_s \ , \nonumber\\
\label{pressure}
\end{eqnarray}
where the free terms  containing the quark momenta and effective masses are:
\begin{eqnarray}
\theta_f=-\frac{i}{2}  {\rm tr}  \int  \frac {d^4 p}{(2\pi)^4} \ln \left(-p^2 + M_f^2 \right ) \ ,
\label{theta} 
\end{eqnarray}
while the scalar condensates are:
\begin{eqnarray}
\phi_f= \langle {\bar \psi}_f \psi_f \rangle= -i  \int \frac {d^4 p}{(2\pi)^4} 
{\rm tr}\frac{1}{(\not \! p - M_f+i\epsilon)} \ ,
\label{phi}
\end{eqnarray}
with traces taken over three colors and Dirac space (but not flavors). The quark effective masses 
can be obtained 
self consistently  from: 
\begin{eqnarray}
 M_i=m_i - 4 G \phi_i + 2K \phi_j \phi_k \ ,  \label{gap_eqs}
\end{eqnarray}
with $(i,j,k)$ being any permutation of flavors $(u,d,s)$. 

We can rewrite Eq.~(\ref{theta}) in terms of a vacuum, a medium and a magnetic contribution
\cite{Menezes:2008qt,Menezes:2009uc}:
\begin{eqnarray}
\theta_f= \theta^{vac}_f+\theta^{med}_f + \theta^{mag}_f \ ,
\end{eqnarray}
which at zero temperature are given by:
\begin{eqnarray}
\theta^{vac}_{f}=- \frac{N_c }{8\pi^2} \left \{ M_f^4 \ln \left [
    \frac{(\Lambda+ \epsilon_\Lambda)}{M_f} \right ]
 - \epsilon_\Lambda \, \Lambda\left(\Lambda^2 +  \epsilon_\Lambda^2 \right ) \right \} \ , \nonumber \\
\end{eqnarray}
where $\epsilon_\Lambda=\sqrt{\Lambda^2 + M_f^2}$,
\begin{eqnarray}
\theta^{med}_f&=&\sum_{\nu=0}^{\nu_{f,max}} \alpha_\nu\frac {|q_f| B N_c }{4 \pi^2}  
\Bigg{[} \mu_f \sqrt{\mu_f^2 - s_f(\nu,B)^2} \nonumber\\
&-&s_f(\nu,B)^2 \ln  \Bigg{(} \frac { \mu_f +\sqrt{\mu_f^2 -s_f(\nu,B)^2}} {s_f(\nu,B)}  \Bigg{)} \Bigg{]}\,,
\nonumber\\
\end{eqnarray}
and
 \begin{eqnarray}
\theta^{mag}_f= \frac {N_c (|q_f| B)^2}{2 \pi^2} \left [ \zeta^{\prime}(-1,x_f) -  
\frac {1}{2}( x_f^2 - x_f) \ln x_f +\frac {x_f^2}{4} \right ],\nonumber \\
\end{eqnarray}
where we have  $x_f = M_f^2/(2 |q_f| B)$ and $\zeta^{\prime}(-1,x_f)= d \zeta(z,x_f)/dz|_{z=-1}$, 
with $\zeta(z,x_f)$  being the Riemann-Hurwitz
zeta function.

We can also rewrite Eq.~(\ref{phi}) in terms of a vacuum, a medium and a magnetic contribution 
\cite{Menezes:2008qt,Menezes:2009uc}, where the four-dimensional integrals gave rise to the sum over Landau levels in the medium contribution:
\begin{eqnarray}
\phi_f=\phi_f^{vac}+\phi_f^{med}+\phi_f^{mag} \ ,
\end{eqnarray}
which at zero temperature are given by:
\begin{eqnarray}
\phi_f^{vac} &=& -\frac{ M_f N_c }{2\pi^2} \left [
\Lambda \epsilon_\Lambda -  {M_f^2} \ln \left ( \frac{\Lambda+ \epsilon_\Lambda}{{M_f }} \right ) \right ] \ ,
\end{eqnarray}
\begin{eqnarray}
\phi_f^{med}&=&
\sum_{\nu=0}^{\nu_{f,max}} \alpha_\nu \frac{ M_f |q_f| B N_c }{2 \pi^2}\nonumber\\
&\times&\left [\ln \left ( \frac { \mu_ f +\sqrt{\mu_f^2 -
s_f(\nu,B)^2}} {s_f(\nu,B)} \right ) \right ]\ ,
\end{eqnarray}
\begin{eqnarray}
\phi_f^{mag}
&=& -\frac{ M_f |q_f| B N_c }{2\pi^2} \Bigg{[} \ln \Gamma(x_f)  \nonumber\\
&-&\frac {1}{2} \ln (2\pi) +x_f -\frac{1}{2} \left ( 2 x_f-1 \right )\ln (x_f) \Bigg{]} \ .
\end{eqnarray}

In the direction perpendicular to the magnetic field, the pressure receives an extra 
contribution due to the 
quantization of the charged particles into the Landau levels \cite{PhysRevC.82.065802,Strickland:2012vu}:
\begin{eqnarray}
P_{\perp}= P + {\cal M} B\  ,
\end{eqnarray}
where the magnetization ${\cal M}$ is going to be calculated in the following section.

For charge neutral $\beta$-equilibrated matter, an additional contribution to the pressure
due to the leptons (electrons and muons) is added:
 \begin{eqnarray}
P_{l} &=& \theta^{med}_{l} = \sum_{\nu=0}^{\nu_{f,max}} \alpha_\nu\frac {B}{4 \pi^2}  
\Bigg{[} \mu_l \sqrt{\mu_l^2 - s_l(\nu,B)^2} \nonumber\\
&-&s_l(\nu,B)^2 \ln  \Bigg{(} \frac { \mu_l +\sqrt{\mu_l^2 -s_l(\nu,B)^2}} {s_l(\nu,B)}  \Bigg{)} \Bigg{]} 
\  ,
\end{eqnarray}
where in the latter expression $s_l=\sqrt{m_l^2+2eB\nu}$. Note that only medium terms contribute to the pressure when a degenerate free gas of leptons is considered.

\subsection{Vector interaction}

One of the options for including a vector interaction in the NJL model gives the following extra term to be 
added to the Lagrangian \cite{Kitazawa:2002bc,Fukushima:2008wg,Bratovic:2012qs,Shao:2013toa,Sasaki:2013mha,Masuda:2012ed}: 
\begin{equation} 
{\pazocal L}_{\rm{vec}}  = - G_V ({\bar \psi} \gamma^\mu \psi)^2 \ ,
\end{equation} 
which was chosen because it reproduces a stiffer equation of state (EoS) and, consequently, more massive 
compacts stars \cite{Menezes:2014aka}. 
The constant $G_V$ was chosen to be equal to $G$ in order to maximize the effects of 
the vector interaction and 
allow one to test their effects in the presence of magnetic fields.

The pressure receives the following extra term due to the chosen vector interaction:
\begin{eqnarray} 
P_{\rm{vec}} =G_V (\rho_u+\rho_d+\rho_s)^2  \label{pre_vec}  \ , 
\end{eqnarray} 
where $\rho=\rho_u+\rho_d+\rho_s=3\rho_B$ is the total quark density. As explained in 
Ref.~\cite{Vogl:1991qt}, 
in the mean field approximation, the role of the vector interaction is to introduce a shift 
in the quark chemical 
potential $\mu_f$ producing  an  effective quark chemical potential, $\tilde{\mu}_f = \mu_f - 2 G_V \rho$
to be taken into account in all thermodynamical quantities. 
\section{Magnetization}
In this section, we  discuss the calculation of the magnetization emphasizing some fundamental 
points, which have been inappropriately considered in some recent 
publications \cite{Menezes:2015mqa,Menezes:2015jka}.  
As explained in detail in Refs.~\cite{LandauLifshitzPitaevskii, PhysRevLett.84.5261,Martinez:2003dz,PhysRevC.82.065802, Strickland:2012vu,1407.2280}, the magnetization 
can be calculated simply by taking the partial derivative of the parallel pressure
or, equivalently, minus the thermodynamical potential with respect to B:
\begin{equation}
 {\cal M}=\frac{\partial P}{\partial B} = -\frac{\partial \Omega }{\partial B} \ .
 \label{magcalc}
\end{equation}

Next, we consider the magnetization calculation for the SU(3) NJL quark model including a
vector interaction, since 
the extension to the simpler cases SU(2) and $G_V$=0 are trivial.
As discussed in the previous section, the introduction of a vector interaction in the
SU(3) NJL model within the mean field approximation is achieved by replacing
the quark chemical potential $\mu_f$  by the corresponding quark effective chemical potential 
$\tilde{\mu}_f = \mu_f - 2 G_{V} \rho $. Thus, the grand potential for the quarks $\Omega(T,\{\mu_f \},B; \{\phi_f\},\rho)$ can be rewritten using 
Eq.~(\ref{pre_vec}):
\begin{eqnarray}
 \Omega&=& \sum_{f=u,d,s} \Omega_f +2G(\phi_u^2+\phi_d^2+\phi_s^2) - 4K \phi_u \phi_d \phi_s \nonumber \\
 &+& G_V \rho^2  , \nonumber \\\label{omega1}
\end{eqnarray}
with each flavor contribution $\Omega_f ( T,M_f,\tilde{\mu}_f ,B)$ given by:
\begin{eqnarray}
&&\Omega_f  = -  \left[ \theta_f^{vac}  +  \theta_f^{med}  \right.  + \left. \theta_f^{mag} \right] \ .
 \label{omega2}
\end{eqnarray}

The thermodynamically consistent solutions (discussed in Ref.~\cite{Buballa:2003qv}) 
correspond to the stationary solutions
of $\Omega$ as a function of $\phi_f$ and $\rho$ and are the key for the correct 
calculation of the magnetization, as shown in the following. It is easy to verify that one of the stationary solutions of the grand potential, Eq.~(\ref{omega1}), is:
\begin{equation}
 \frac{\partial \Omega}{\partial \phi_f} =0 \ , \label{constgap}
\end{equation}
 where the condensate is given by:
\begin{equation}
  \phi_f=\frac{\partial \Omega  }{\partial  m_f} = \frac{\partial \Omega  }{\partial  M_f} \ ,  
\end{equation}
with the corresponding gap equation, Eq.~(\ref{gap_eqs}).
 Due to presence of the vector 
interaction, the grand potential becomes an explicit function of the total quark density and,
once more, in order to assure 
thermodynamical consistency \cite{Vogl:1991qt,Buballa:2003qv}, we have to impose a second 
stationary condition:
\begin{eqnarray}
 \frac{\partial \Omega}{\partial \rho} =  0 = \sum_{f=u,d,s} \frac {\partial \Omega}{\partial {\tilde \mu_f}}  
  \frac{\partial {\tilde \mu_f} }{\partial \rho} +2 G_V \rho \ ,
\end{eqnarray}
where we have used Eq.~(\ref{omega1}) and:
\begin{equation}
   \rho_f= - \frac{\partial\Omega}{\partial \mu_f} = -
                    \frac{\partial\Omega}{\partial \tilde{\mu}_f} \ .
                    \label{constrho}
\end{equation}
The constraint in the latter equation simply means that the total quark density 
has to satisfy the condition: $\rho=\rho_u+\rho_d+\rho_s$ in equilibrium. Both the gap equation and 
the latter constraint have
to be simultaneously and self-consistently solved.

From Eq.~(\ref{magcalc}), one may write:
\begin{eqnarray}
{\cal M}=-\left. \frac{\partial \Omega}{\partial B}
\right|_{\{ \phi_f \},\rho}  - \frac{\partial \Omega}{\partial \phi_f} 
\frac{\partial \phi_f}{\partial B}
- \frac{\partial \Omega}{\partial \rho}
\frac{\partial \rho}{\partial B} \ ,
\end{eqnarray} 
which can be simplified using Eq.~(\ref{omega2}) and the constraints 
given by Eqs.~(\ref{constgap} and \ref{constrho}), 
yielding the following expression:
{\small
\begin{equation}
 {\cal M}=  \sum_{f} \left[ \frac{\partial  \theta_f^{med} }{\partial B} +
 \frac{\partial \theta_f^{mag}}{\partial B} \right] \ . \label{magcorr}
\end{equation}   }
This expression shows that only two terms contribute to the magnetization.  
Note that in Refs.~\cite{Menezes:2015mqa,Menezes:2015jka} the derivatives of the $\phi's$ were 
incorrectly
taken as being non-zero. Hence, a spurious increase of orders of magnitude 
was found in the magnetization in these references, which 
generated incorrect results for the perpendicular pressure leading 
the authors to 
erroneously conclude that strong anisotropy effects could exist for magnetic fields as small as 
10$^{17}$ Gauss. 

The (non-zero) derivatives $\theta'$ at $T=0$ are as in Ref.~\cite{Menezes:2015jka}
\begin{eqnarray}
\theta^{\prime \, med}_f&=& \frac{\theta_f^{med}}{B}- \frac{N_c B |q_f|^2}{2\pi^2} \sum_{\nu=0}^{\nu_{max}} 
\alpha_\nu \nonumber \\  &\times& \ln \left ( \frac{ \tilde{\mu}_f + \sqrt{\tilde{\mu}_f^2-s_f^2}}{s_f} \right ) \nu \ ,
\label{mag1}
\end{eqnarray}
\begin{eqnarray}
\theta^{\prime \, mag}_f&=& 2 \frac{\theta^{mag}_f}{B} -
 \frac{N_c |q_f| M_f^2}{4\pi ^2}  \Bigg{[} \ln
 \Gamma(x_f)-\frac{1}{2} \ln (2 \pi) \nonumber\\
 &+& x_f - (x_f - \frac{1}{2}) \ln(x_f) \Bigg{]} \ .
 \label{mag2}
 \end{eqnarray}
Note that, for the calculation within the 
SU(2) NJL model, we have only to take the summation over the flavors $u$ and $d$ in magnetization 
expression, Eq.~(\ref{magcorr}).

For charge neutral $\beta$-equilibrated matter, the presence of the free gas
of leptons gives an additional contribution to the 
magnetization, $\theta^{\prime}_l$, 
\begin{eqnarray}
\theta^{\prime \, med}_{l} &=& 
\frac{\theta_l^{med}}{B}- \frac{B}{2\pi^2} \sum_{\nu=0}^{\nu_{max}} 
\alpha_\nu \nu \ln \left ( \frac{ \mu_l + \sqrt{\mu_l^2-s_l^2}}{s_l} \right )  \ , \nonumber\\
\label{maglep}
\end{eqnarray}
where only the medium term appears.

\section{Results and discussion}

\begin{figure}[!]
\includegraphics[width=7.5cm]{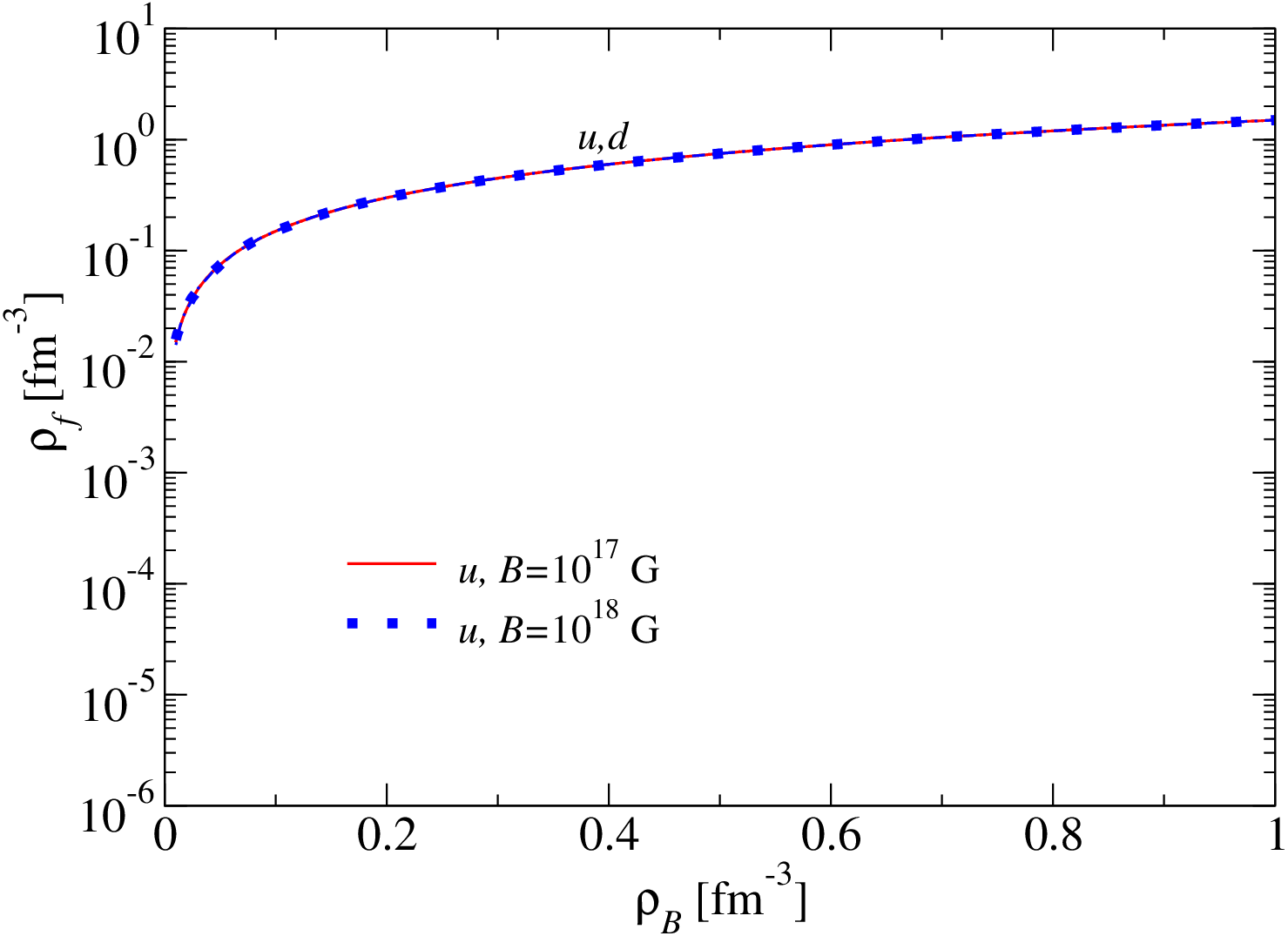}
\caption{(Color online) Number density of each quark flavor for zero-strangeness isospin-symmetric matter. 
The different curves show density-independent magnetic field strengths of $10^{17}$ G and  $10^{18}$ G.
All curves overlap.}
\end{figure}

\begin{figure}[t]
\includegraphics[width=7.5cm]{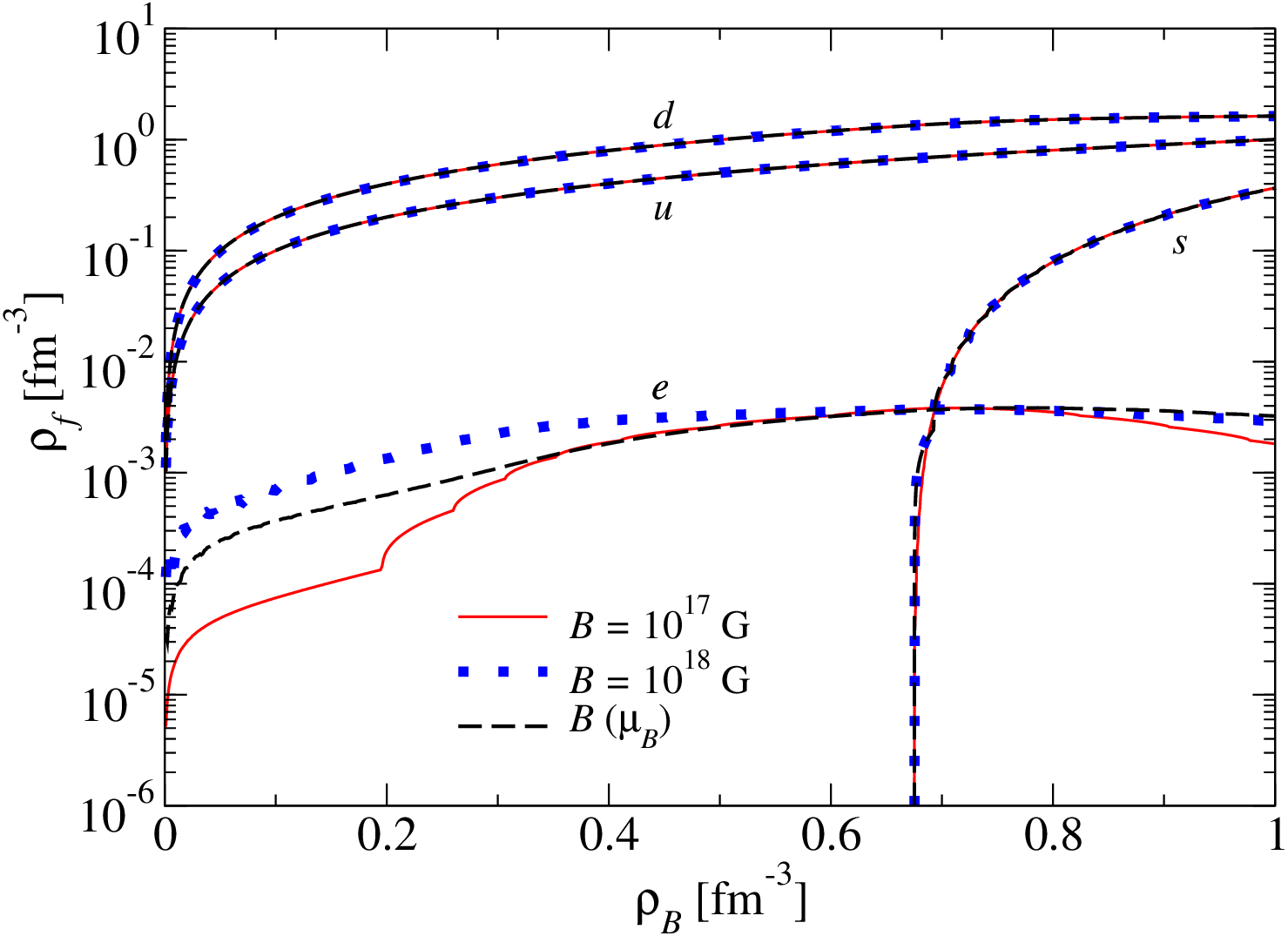}
\caption{(Color online) Same as Fig.~1 but for charge neutral $\beta$-equilibrated neutron-star matter 
with electrons (the muons do not appear) also showing a realistic stellar chemical potential-dependent magnetic field 
from Ref.~\cite{Dexheimer:2016yqu}.}
\end{figure}

To exemplify the differences that appear when taking the derivatives of the $\phi's$ as (correctly) being zero, 
we remake some of the figures from  Refs.~\cite{Menezes:2015mqa,Menezes:2015jka} and point out the differences we find. In 
each figure we show different quantities as a function of baryon number density $\rho_B=\sum_f \rho_f/3$.  

In different figures we show results for zero-strangeness isospin-symmetric matter and charge-neutral $\beta$-equilibrated neutron-star matter with leptons. In the first case, zero-strangeness is enforced at zero temperature simply by not including the strange quark. Isospin-symmetric matter is enforced by using the same chemical potential for up and down quarks $\mu_u=\mu_d=\mu_B/3$, where $\mu_B$ is the baryon chemical potential. In the second case, charge neutral matter is enforced by 
\begin{equation}
\sum_i q_i \rho_i=0\ ,
\end{equation}
where the index $i$ runs over quarks and leptons. $\beta$-equilibrium allows us to rewrite the fermion chemical potentials as a function of the chemical potentials related to the conserved quantities in the system, baryon and charge chemical potentials
\begin{equation}
\mu_u = \frac 1 3 \mu_B + \frac 2 3 \mu_q\ ,
\end{equation}
\begin{equation}
\mu_d = \frac 1 3 \mu_B - \frac 1 3 \mu_q\ ,
\end{equation}
\begin{equation}
\mu_s= \frac 1 3 \mu_B - \frac 1 3 \mu_q\ ,
\end{equation}
\begin{equation}
\mu_e= \mu_\mu = - \mu_q\ ,
\end{equation}

In different figures we show results for a density-independent magnetic field of strength $10^{17}$ G, a density-independent magnetic field of strength $10^{18}$ G, and a realistic (polar) stellar chemical potential-dependent field \cite{Dexheimer:2016yqu}
\begin{eqnarray}
B^*(\mu_B)=\frac{(a + b \mu_B + c \mu_B^2)}{B_c^2} \ \mu,
\label{3}
\end{eqnarray}
with the baryon chemical potential $\mu_B$ given in MeV and the dipole magnetic moment $\mu$ in Am$^2$ in order 
to produce $B^*$ in units of the critical field of the electron $B_c=4.414\times 10^{13}$ G. The value of the 
coefficients for a star with baryon mass $M_B=1.6$ M$_\odot$(that gives a gravitational mass 
$\sim1.4$ M$_\odot$) are $a=-1.02$ G$^2$/(A m$^2$), $b=1.58\times10^{-3}$ G$^2$/(A m$^2$ MeV), and
$c=-4.85\times10^{-7}$ G$^2$/(A m$^2$ MeV$^2$). We choose $\mu=2\times10^{32}$ Am$^2$, which 
reproduces field strengths between $1.03\times10^{17}$ and $5.31\times10^{17}$ G in a $M_B=1.6$ M$_\odot$ star
in the presence of a vector interaction.

\begin{figure}[t!]
\includegraphics[width=7.5cm]{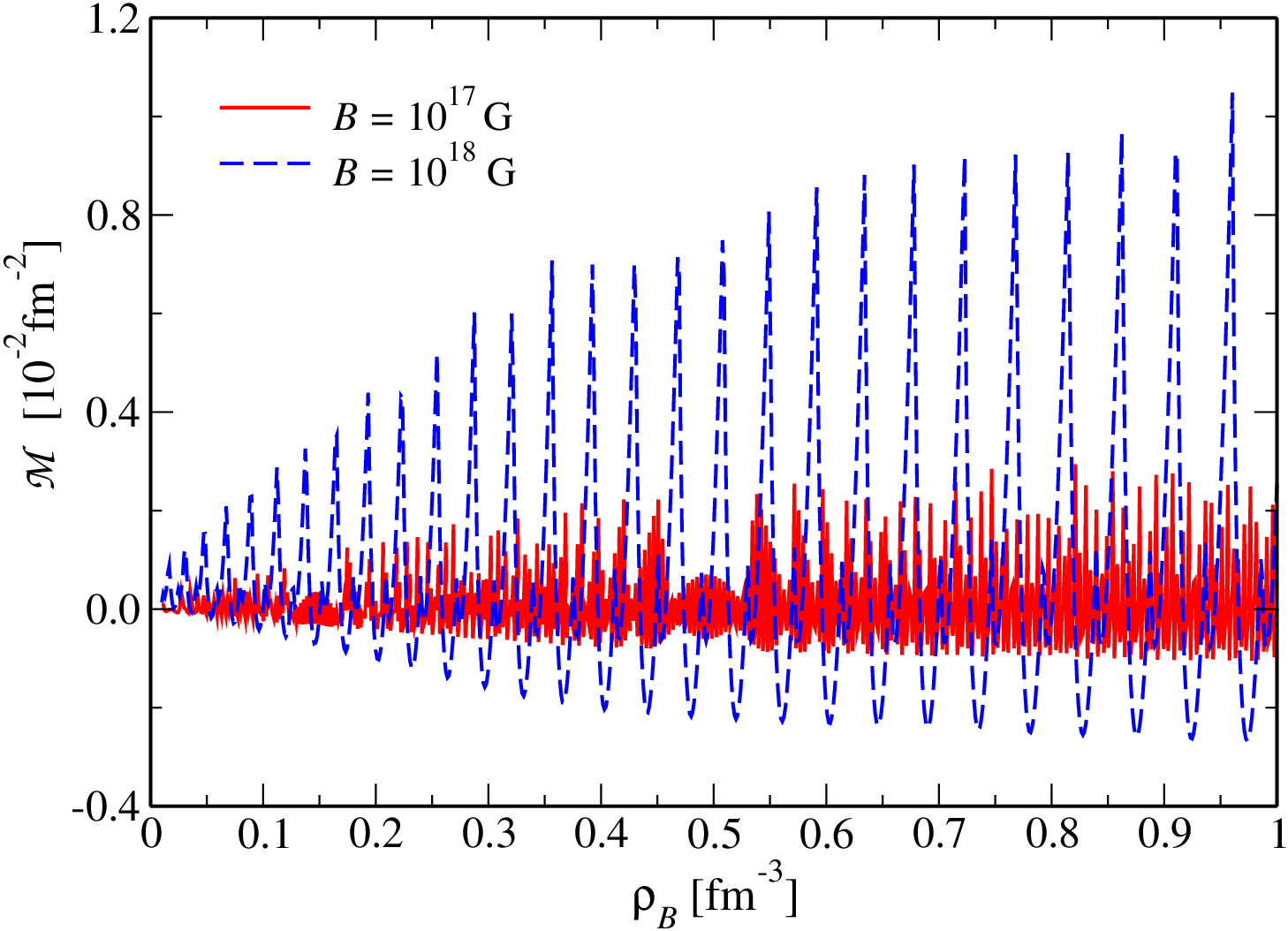}
\caption{(Color online) Magnetization for zero-strangeness isospin-symmetric matter. The different curves show density-independent magnetic field strengths of $10^{17}$ G and  $10^{18}$ G.}
\end{figure}

Figures 1 and 2 show the number density of each quark flavor (and leptons) for isospin-symmetric matter and 
neutron-star matter. In both cases, the effect of realistic magnetic fields (we did not include the case of a 
density-independent magnetic field with $10^{19}$ G) practically cannot be seen in the plots, in agreement with Refs.~\cite{Menezes:2015mqa,Menezes:2015jka}. 
As expected, in the zero-strangeness isospin-symmetric matter case, the amount of up and down quarks is the same 
for any baryon number density.

Figures 3 and 4 show the magnetization of the system for isospin-symmetric matter and neutron-star matter. 
In both cases, the effect of realistic magnetic fields produce magnetizations about three orders of magnitude lower in strength 
than in Refs.~\cite{Menezes:2015mqa,Menezes:2015jka} and that oscillate (as in Ref.~\cite{Huang:2009ue}) 
but much less than in Refs.~\cite{Menezes:2015mqa,Menezes:2015jka}. The oscillations in the magnetization are unavoidable 
as Eqs.~(\ref{mag1}) and (\ref{mag2}) have positive and negative terms. Note that even a free-Fermi gas produces 
oscillations in the magnetization with positive and negative values (for large enough magnetic field strength and
density, see Refs.~\cite{Strickland:2012vu} and references therein for details). Note that 
the magnetization for the chemical-potential-dependent profile in Fig.~4 lies between the ones 
for fixed values $B=10^{17}$ and $B=10^{18}$ G.

\begin{figure}[t!]
\includegraphics[width=7.5cm]{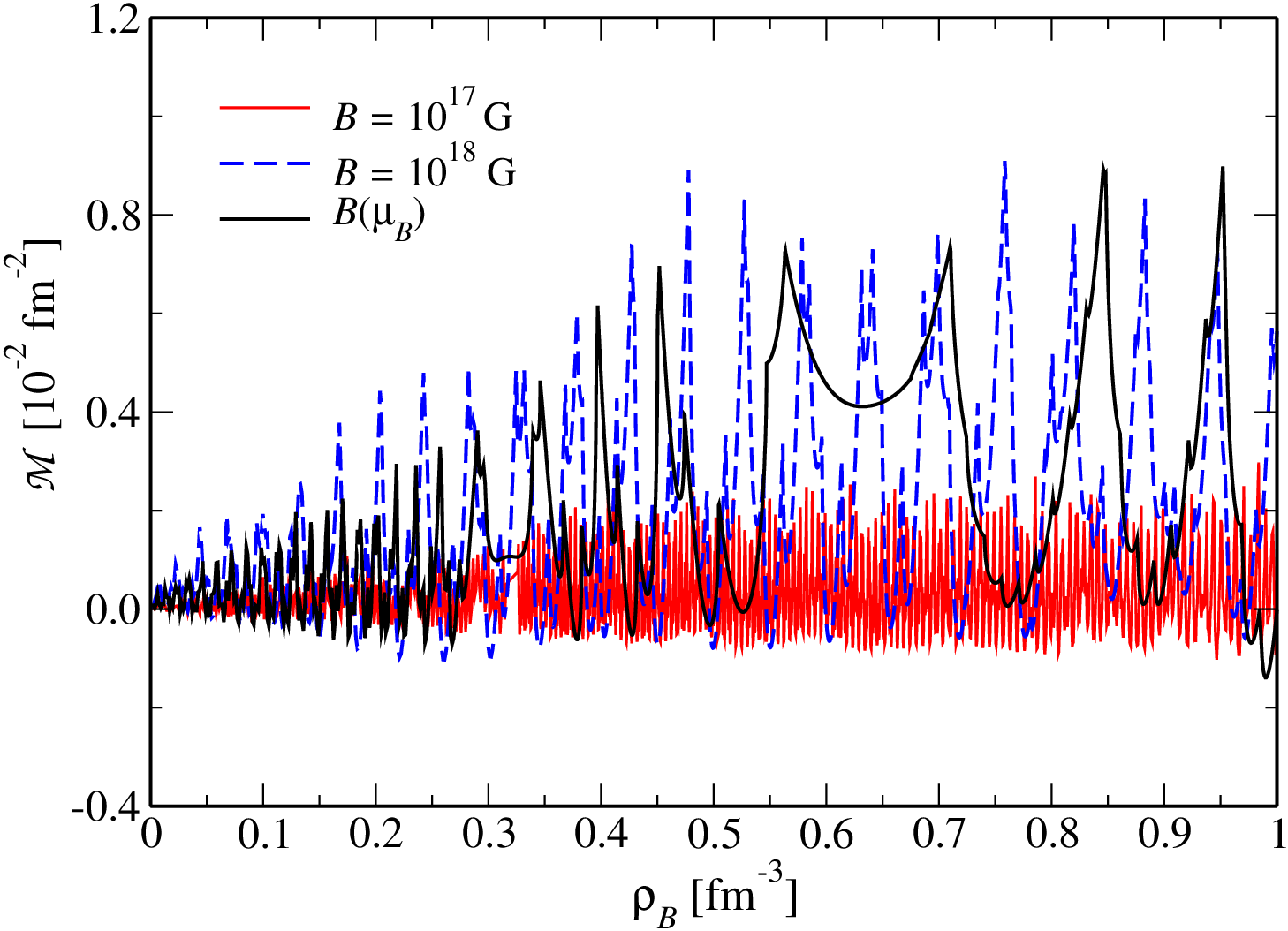}
\caption{(Color online) Same as Fig.~3 but for charge neutral $\beta$ equilibrated neutron-star matter with electrons also showing a realistic stellar chemical potential-dependent magnetic field 
from Ref.~\cite{Dexheimer:2016yqu}.}
\end{figure}

Figures 5 and 6 show the parallel pressure of the system for isospin-symmetric matter 
and neutron-star matter.  In both cases, the effect of realistic magnetic fields cannot be seen in the plots.  
This is in agreement with results from Refs.~\cite{Menezes:2015mqa,Menezes:2015jka}, except for the case with density-independent 
magnetic field with strength of $10^{19}$ G (not shown in our plots), in which case we would see the unphysical
behavior of pressure going up and down with the increase of baryon number density or energy density, which means that the NJL model is unstable under those unphysical conditions. The reduction in the increase 
of pressure around $0.7$ fm$^{-3}$ in Fig.~6 is related to the appearance of the strange quarks in the system. 
The negative pressures at low densities in Figs.~5 and 6, on the other hand, indicate the presence of coexisting phases and associated phase transitions at those densities or, in other words, a crust is required for star stability \cite{Hanauske:2001nc}.

\begin{figure}[t!]
\includegraphics[width=7.5cm]{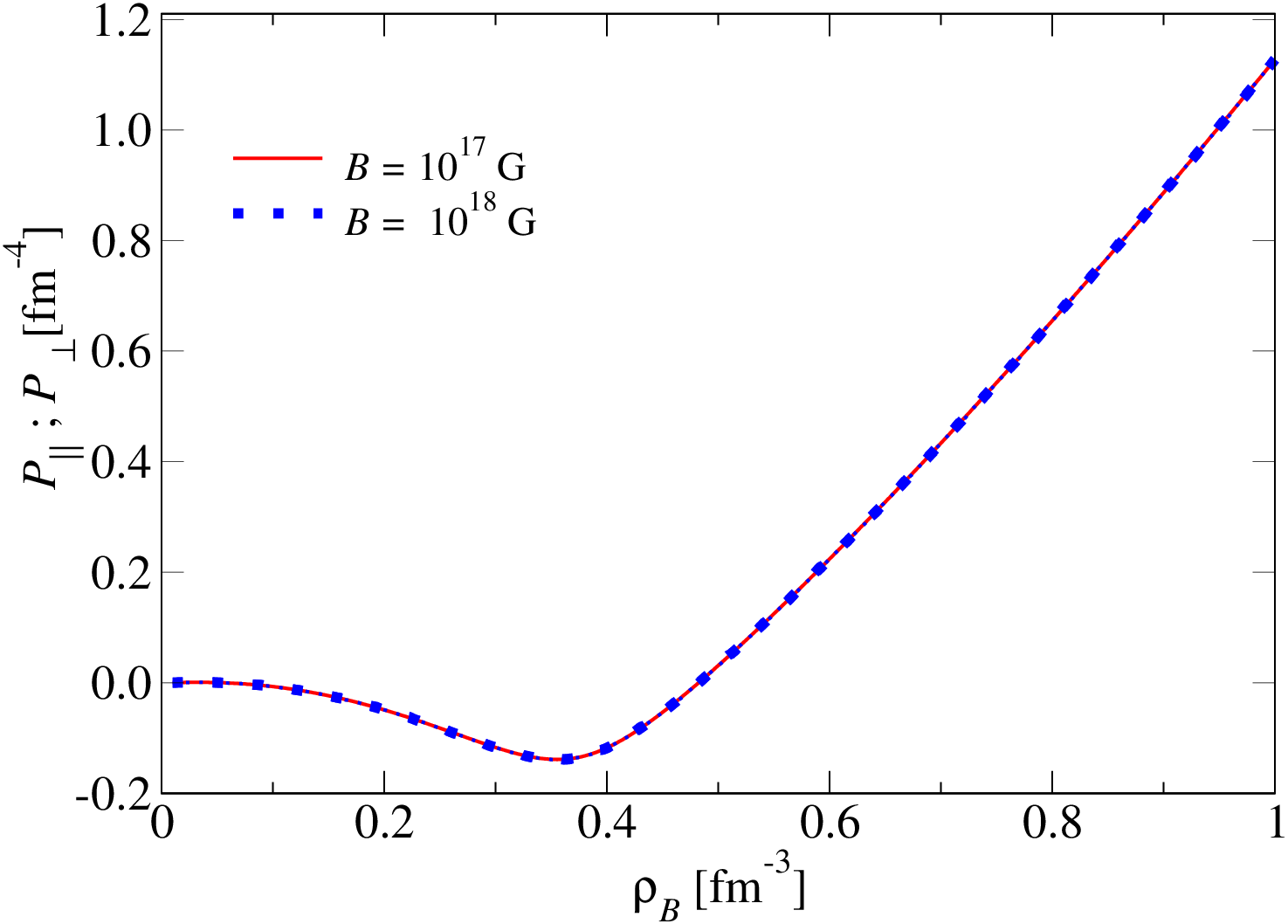}
\caption{(Color online) Parallel and perpendicular pressures for zero-strangeness isospin-symmetric matter. The different curves show density-independent magnetic field strengths of $10^{17}$ G and  $10^{18}$ G. All curves overlap.}
\end{figure}

\begin{figure}[t!]
\includegraphics[width=7.5cm]{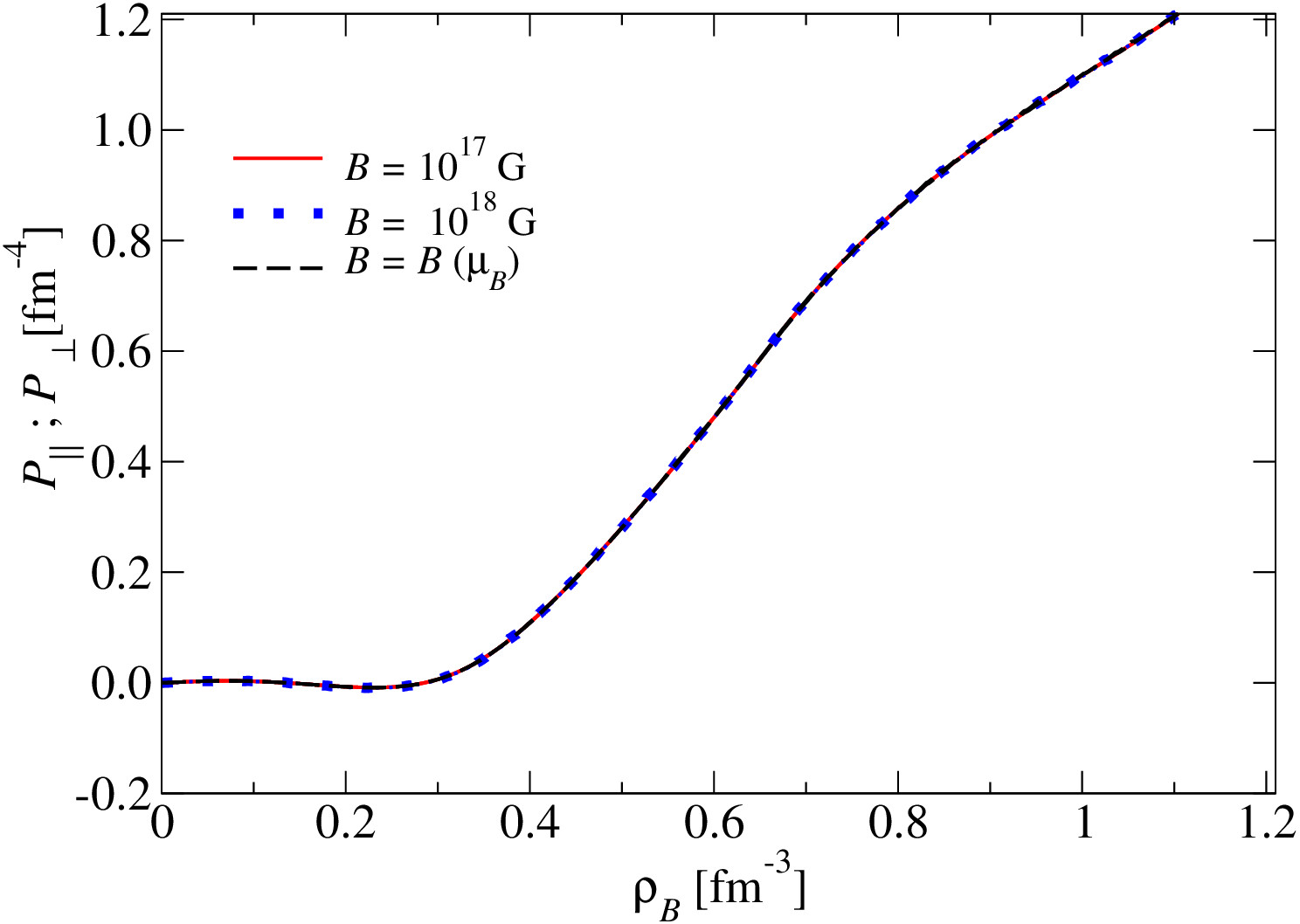}
\caption{(Color online) Same as Fig.~5 but for charge neutral $\beta$ equilibrated neutron-star matter with electrons also showing a realistic stellar chemical potential-dependent magnetic field 
from Ref.~\cite{Dexheimer:2016yqu}. All curves overlap.}
\end{figure}

The perpendicular pressure of the system for isospin-symmetric matter and neutron-star matter is almost equal to the respective parallel pressures (Figs.~5 and 6), when using realistic magnetic fields (as already shown in Ref.~\cite{Dexheimer:2012mk} using the bag model).  This is in disagreement with results from Refs.~\cite{Menezes:2015mqa,Menezes:2015jka}, in which case the perpendicular pressures are 
different for different magnetic field strengths, different from the respective parallel pressures 
and, most importantly, discontinuous. 

Figure 7 shows that for magnetic field strengths slightly larger than of $10^{18}$ G, the 
pressure in the direction of the magnetic field (parallel) and perpendicular to it start to be different at 
any baryon number density. As already discussed in detail in Refs.~\cite{Chatterjee:2014qsa,Franzon:2015sya,Dexheimer:2016yqu}, 
self-consistent general-relativity calculations assuming poloidal magnetic fields do not present solutions for 
stars that posses central magnetic fields beyond $1$ or $2$ times $10^{18}$ G. In this case, there would be no difference in
using an EoS with magnetic field effects as input for those calculations (unlike what was stated in Refs.~\cite{Menezes:2015mqa,Menezes:2015jka}). Note, however, that there are no 
consistent general relativity calculations of such kind using the NJL model with a vector interaction, in which case a much stiffer EoS might 
allow larger stellar central magnetic fields. This issue will be addressed in a future publication.

\begin{figure}[t!]
\includegraphics[width=7.5cm]{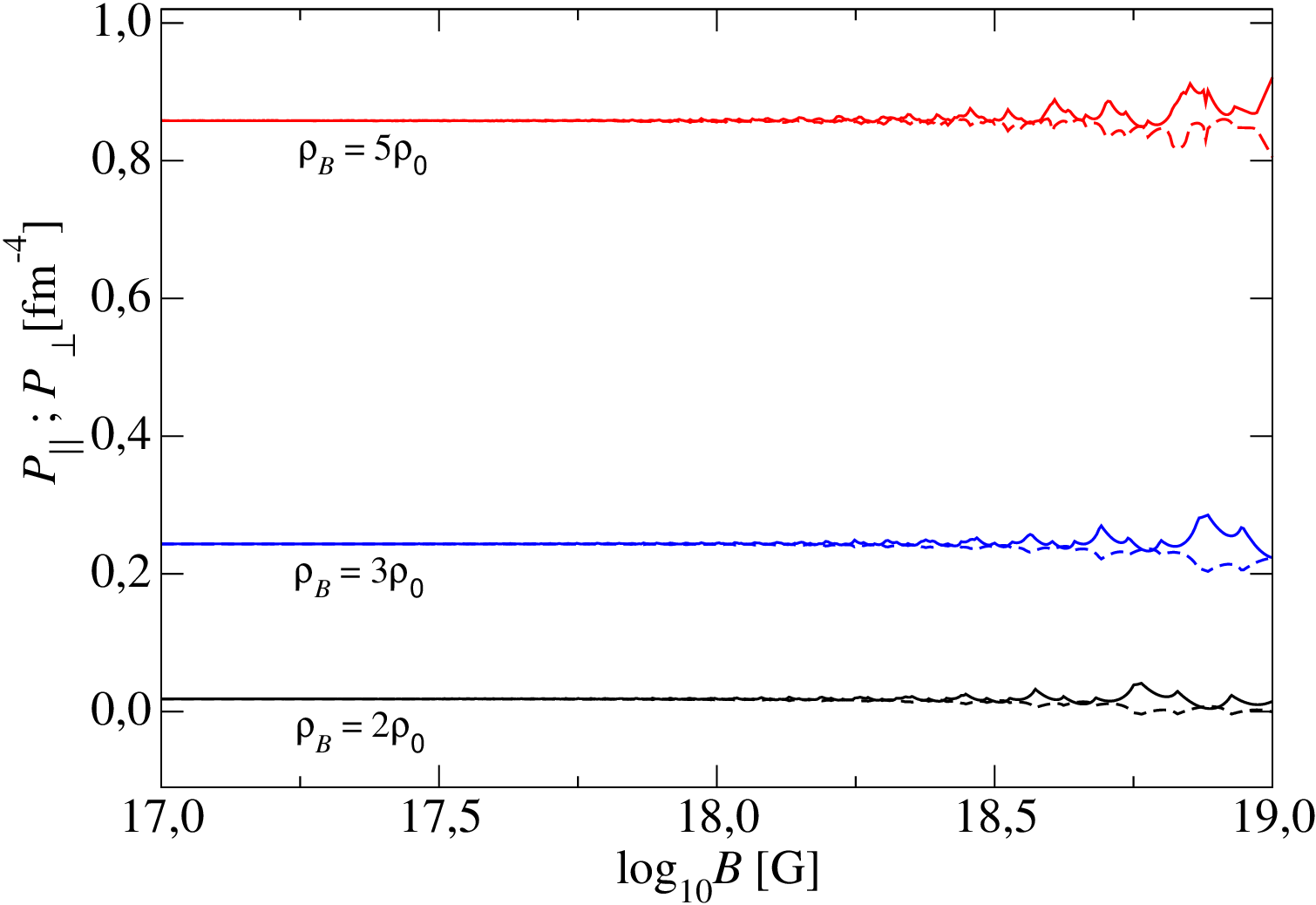}
\caption{(Color online) Parallel (solid lines) and perpendicular (dotted lines) pressures for charge neutral 
$\beta$ equilibrated neutron-star matter with electrons as a function of magnetic field strength. The different 
curves show the pressures at different baryon number densities.}
\end{figure}

\begin{figure}[t]
\includegraphics[width=7.5cm]{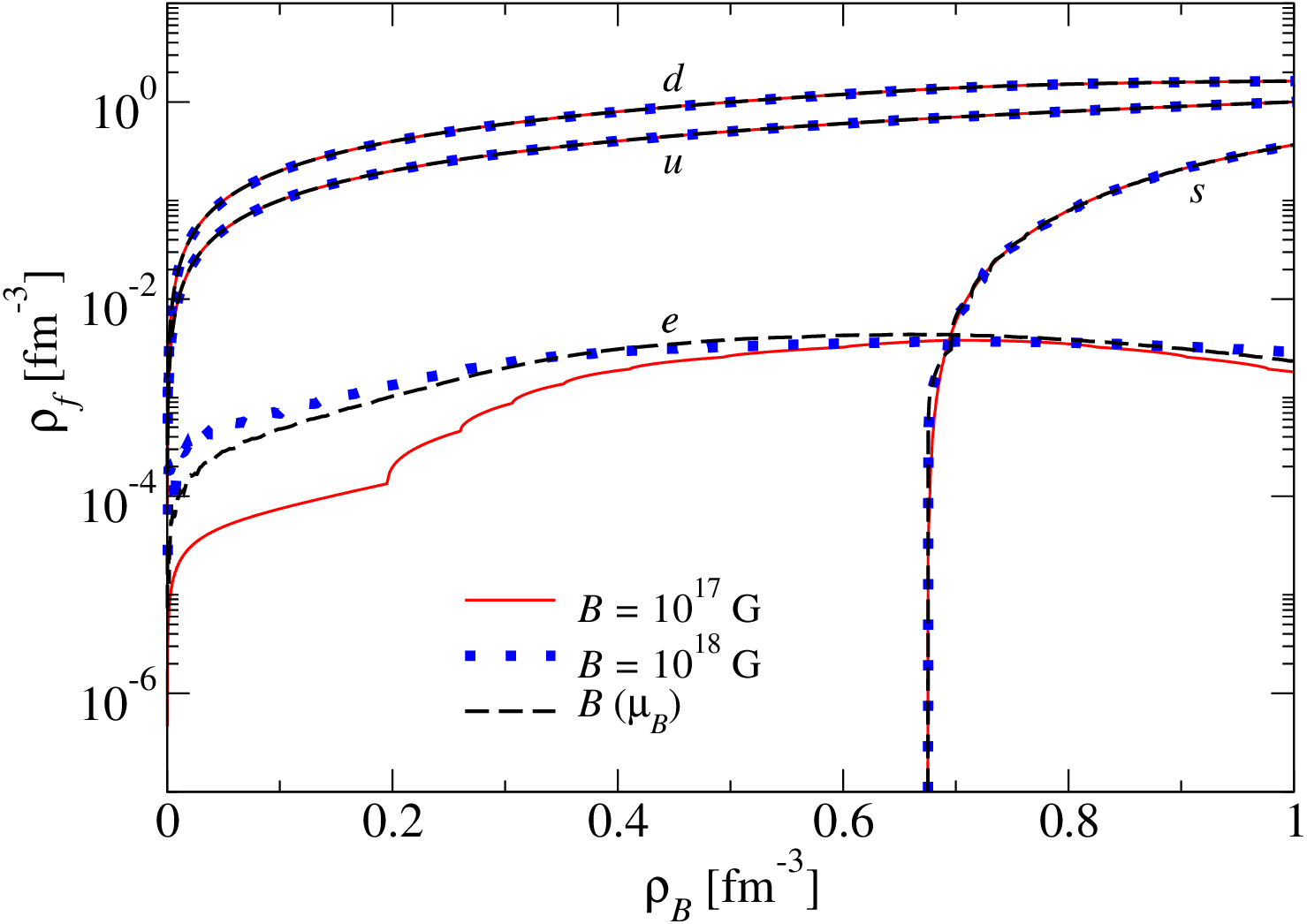}
\caption{(Color online) Same as Fig.~2 but including a vector interaction.}
\end{figure}

Next, we present some results for charge neutral $\beta$-equilibrated matter including a vector interaction. These results
are of special importance for magnetars and to our knowledge the magnetization study in this case
has not yet been done in the literature.
The change in the population due to the inclusion of a vector interaction is very small as can be seen 
in Fig.~8 (when compared with Fig.~2). Once more, the effect of all magnetic field strengths analyzed in this 
work practically does not modify the population. The changes in the magnetization with the inclusion of a vector 
interaction, on the other hand, are considerable, as can be seen in Fig.~9 (when compared with Fig.~4). The interaction leads the 
magnetization to oscillate more as a function of baryon number density, although it still has a low magnitude.

\begin{figure}[t!]
\includegraphics[width=7.5cm]{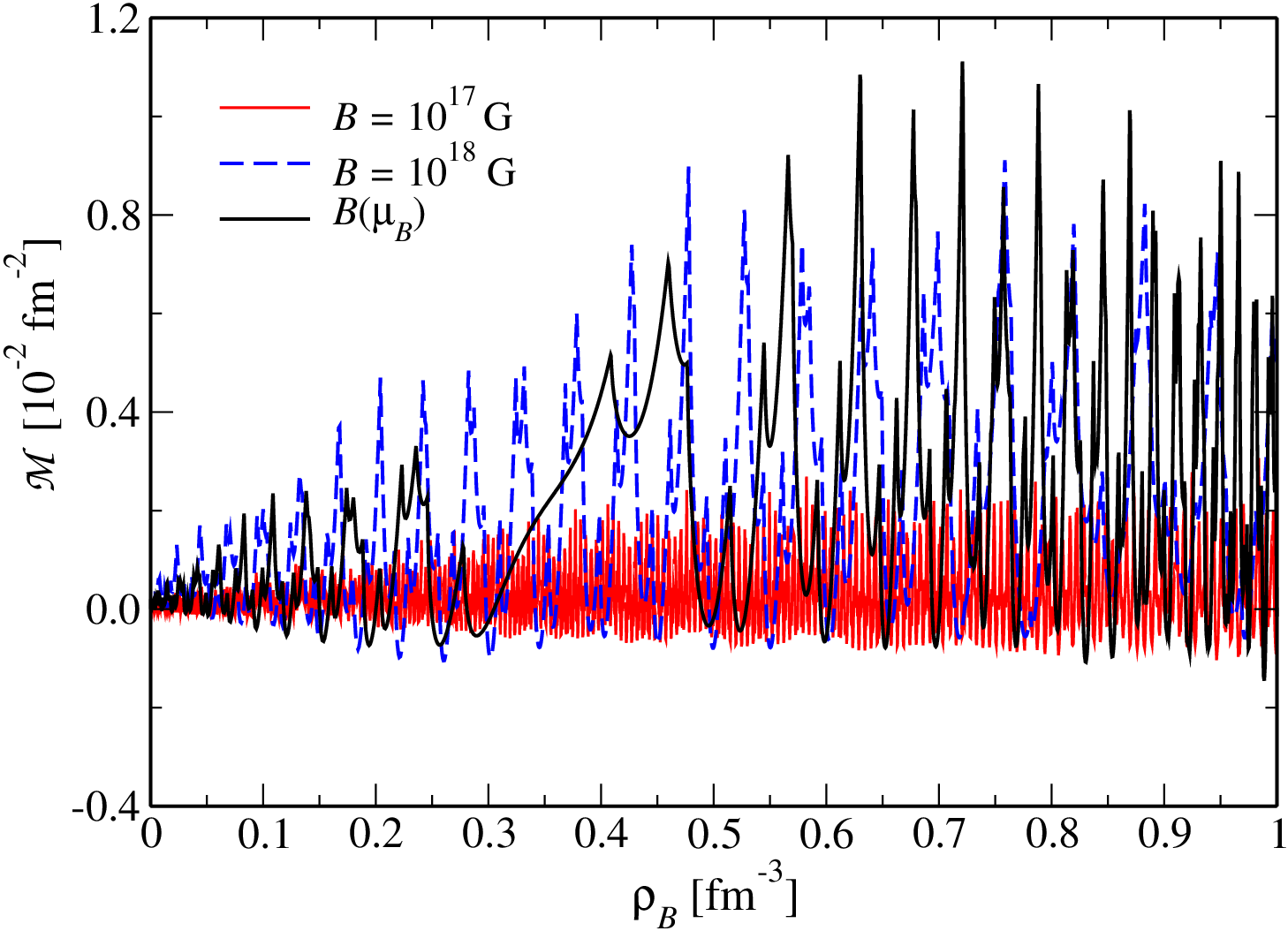}
\caption{(Color online) Same as Fig.~4 but including a vector interaction.}
\end{figure}

\begin{figure}[t!]
\includegraphics[width=7.5cm]{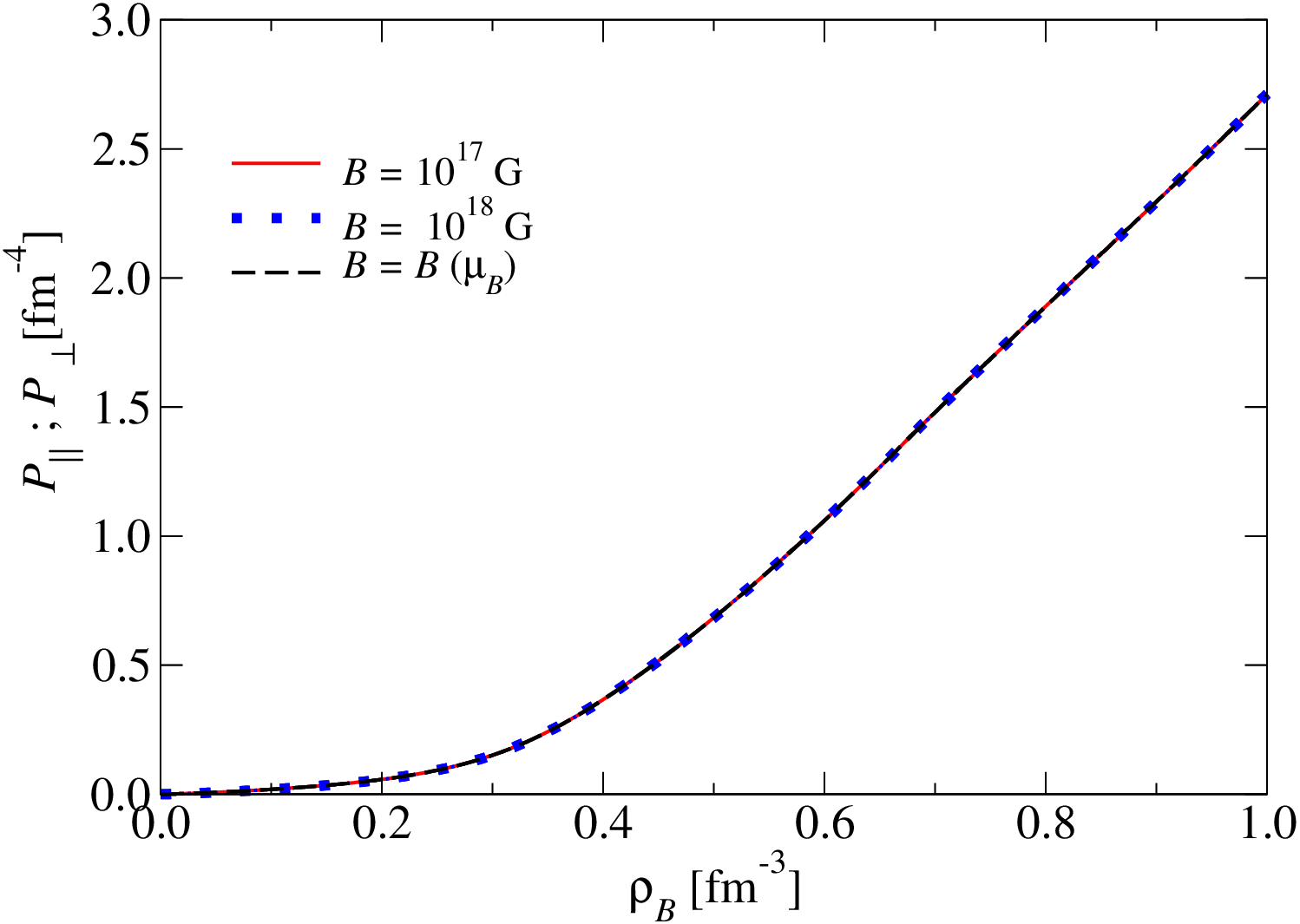}
\caption{(Color online) Same as Fig.~6 but including a vector interaction.}
\end{figure}

As already discussed in the beginning of the section, when calculated correctly, the parallel and perpendicular 
pressures are almost equal when considering realistic magnetic fields. This does not change with the inclusion of a vector
interaction as seen in Fig.~10. When compared with Fig.~6, it can be seen that the interaction makes the
EoS of neutron star matter much stiffer, which is exactly why it is necessary to reproduce massive 
neutron stars. Note however that it has been shown that, for small chemical potentials, zero or a small vector interaction
in quark matter is required in order to be in agreement with lattice QCD and perturbative QCD concerning baryon 
number susceptibilities \cite{Steinheimer:2014kka, Bazavov:2013uja,Haque:2013sja}. Nevertheless, whether or not those constraints
are required at large chemical potentials is still an open question. Finally, we note that our results for the parallel pressures are in agreement with the results from Ref.~\cite{Menezes:2014aka}, which includes magnetic field effects but does not calculate the magnetization.

\section{Conclusions and Outlook}

In this work, we showed that the inclusion of magnetic fields in quark matter within the Nambu-Jona-Lasinio
formalism does not generate the unphysical behavior previously found in Refs.~\cite{Menezes:2015jka,Menezes:2015mqa} 
for the magnetization and pressure in the direction perpendicular to the magnetic field.
We showed this for realistic magnetic fields, for zero-strangeness isospin-symmetric and neutron star 
(chemically equilibrated and charge neutral) matter, including or not a vector interaction.  
We found that up to magnetic field strengths 
of $10^{18}$ G, there is no change in population and there is no pressure anisotropy generated
by the magnetic field (the pure electromagnetic 
contribution is not accounted for due to the normalization of the pressure). Although 
the magnetization oscillates with baryon number density, its magnitude is very small. 
A careful discussion on the calculation of the magnetization is presented and, in particular, 
the case where a vector interaction is present is analyzed, since, to our knowledge, this has not been done 
in the literature and is very important for the study of magnetars.

In a future publication, we are going to include our results with a strong vector interaction in a fully self-consistent 
general-relativity calculation to investigate if the interaction can increase the maximum stable stellar central magnetic field strength. A magnetic field strength of about $3\times10^{18}$ G will already generate effects of the magnetic field in the equation of state.

\section*{Acknowledgments}

The authors acknowledge support from NewCompStar COST Action MP1304 (V.D.) and from the LOEWE program HIC for FAIR (V.D). Work partially financed by CNPq under grants 308828/2013-5 (R.L.S.F), 306484/2016-1 (S.S.A), and  306195/2015-1(V.S.T.) and FAPESP2016/07061-3 (V.S.T.). We thank M. B. Pinto for discussions and useful comments.

\section*{References}

\bibliographystyle{apsrev4-1}
\bibliography{paper}

\end{document}